\documentclass[useAMS]{mn2e}

\usepackage{txfonts}
\usepackage{graphicx}

\usepackage{graphics,epsf}
\usepackage{amssymb}                
\usepackage{epsfig}                 

\def \s{~\rm{s}}
\def \km{~\rm{km}}

\def \K{~\rm{K}}

\def \AU{~\rm{AU}}

\def \yr{~\rm{yr}}

\def \mum{~\rm{\mu m}}

\title[Transient absorption lines in $\eta$ Car]
{Explaining the transient fast blue absorption lines in the massive binary system $\eta$ Carinae}
\author[A. Kashi et al.]{Amit Kashi$^{1}$\thanks{kashia@physics.technion.ac.il},
Noam Soker$^{1}$\thanks{soker@physics.technion.ac.il},
and Muhammad Akashi$^{1}$\thanks{akashi@physics.technion.ac.il}\\
$^{1}$Department of Physics, Technion -- Israel Institute of Technology, Haifa 32000, Israel}
\begin{document}

\date{Accepted 0000; Received 0000; in original form 2010 October 3}

\pagerange{\pageref{firstpage}--\pageref{lastpage}} \pubyear{2010}

\maketitle

\label{firstpage}

\begin{abstract}
We use recent observations of the He~I~$\lambda10830 \rm{\AA}$ absorption line and
3D hydrodynamical numerical simulations of the winds collision,
to strengthen the case for an orientation of the semimajor axis of the massive binary
system $\eta$ Carinae where the secondary star is toward us at periastron passage.
Those observations show that the fast blue absorption component exists for only several
weeks prior to the periastron passage.
We show that the transient nature of the fast blue absorption component supports a geometry where
the fast secondary wind, both pre and post-shock material,
passes in front of the primary star near periastron passage.
\end{abstract}


\section{INTRODUCTION}
\label{sec:intro}

$\eta$ Car is a very massive stellar binary system, with an orbital period of
$5.54 \yr$ (Damineli 1996), as observed in all wavelengths (e.g., radio, Duncan \& White 2003;
IR, Whitelock et al. 2004; visible, van Genderen et al. 2006, Fernandez-Lajus et al. 2009;
UV, Smith et al. 2004; emission and absorption lines, Nielsen et al. 2009,
Damineli et al. 2008a, b; X-ray, Corcoran 2005, 2010, Hamaguchi et al. 2007).
The high eccentricity of $e \simeq 0.9$ results in rapid changes in emission and absorption lines,
as well as in the continuum, near each periastron passage.
The several weeks of the rapid changes occurring every orbital period is termed the spectroscopic event.
These lines might originate in different places in the binary system: the primary star, the secondary
star, their respective winds, and the colliding winds structure which is termed the conical shell.

As $\eta$ Car is the best studied binary luminous blue variable (LBV),
it holds the key to our understanding of other LBVs.
It is particulary important to understand the behavior near periastron passage, where
the strongest binary interaction takes place, and for that is crucial
to know the orientation of the binary system.
Namely, the direction of the primary more massive
LBV star relative to its less massive but hotter companion at periastron passage.
While it is agreed that the inclination of the binary system is $i\simeq 41^\circ$
(Davidson et al. 2001; Smith 2006), there is
no agreement on the direction of the semimajor axis, termed periastron longitude.
The orientation is measured by the angle $\omega$: $\omega=0^\circ$ for the case
when the secondary moves toward us before periastron passage and the semimajor axis is
perpendicular to the line of sight, $\omega=90^\circ$ when the secondary is toward us at periastron passage, and
$\omega=270^\circ$ when the primary is toward us at periastron passage.
Several properties of the binary system have been used to deduce the orbital orientation,
with contradicting results (for details see Kashi \& Soker, 2008b, 2009c).

One of the properties that led to a contradicting conclusion on the orientation is the behavior
of the blue absorption wing of the He~I~$\lambda10830 \rm{\AA}$ line.
In Kashi \& Soker (2009b) we constructed a toy model where the material responsible
for the blue absorption wing was assumed to reside in the colliding winds region -- the
conical shell -- close to the binary system.
This model is able to account for the transient appearance of the blue absorbing wing
and to the finding that the maximum absorbing velocity is reached several
days before periastron passage, only if the secondary is toward us near periastron passage,
i.e., $\omega \simeq 90^\circ$.
In a recent paper Groh et al. (2010; hereafter G2010) reached an opposite conclusion.
Comparing their observations with a model based on 3D numerical simulations
of the colliding winds structure, G2010 suggested that the orientation is that of $\omega=243^\circ$.
Namely, the primary is closer to us just before periastron passage.
In this paper we critically reexamine both models.

\section{PROPERTIES OF THE ABSORPTION PROFILE}
\label{sec:profile}

The He~I~$\lambda10830 \rm{\AA}$ line has been observed for more than a decade,
but only occasionally (Damineli et al. 1998, Groh et al. 2007; Damineli et al. 2008b; G2010).
The line shows a P-Cygni profile varying in time, especially close to the spectroscopic event.
It has a complicated emission with three spikes, and absorbing components with
velocities of up to $-v \simeq 1900 \km \s^{-1}$.
In this section we describe the relevant properties of the blue absorption wing that rapidly appears and
disappears near periastron passages.

\begin{figure}
\includegraphics[width=1\columnwidth]{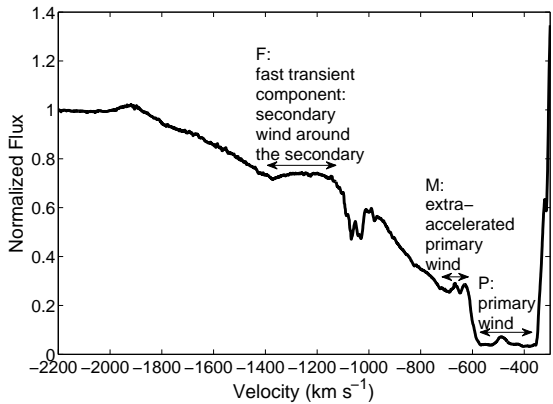}
\caption{
The flat part of each of the three absorption components identified in the paper ,marked on the absorption profile at phase 11.998
(four days before the periastron passage of the 2009 spectroscopic event) taken from G2010.
}
\label{fig:profile}
\end{figure}

G2010 observed the He~I~$\lambda10830 \rm{\AA}$ across the 2009 periastron passage.
In Fig. \ref{fig:profile} we show the absorption profile at phase 11.998,
four days before the 2009 spectroscopic event, assumed to occur at phase 12.000
(Damineli et al. 2008b, Groh \& Damineli 2004),
taken from fig. 1 of G2010.
The profile at phase 11.991 is practically similar in shape,
but the absorption is somewhat weaker.

At phases 11.998 and 11.991 we can identify three flat parts in
the spectrum, as marked on Fig. \ref{fig:profile}.
\begin{enumerate}
\item \emph{P-component.} The deepest component can be safely identified with the regular primary wind.
This part exists during all phases, it is deep, and its velocity matches that of the primary wind.
The primary wind reaches terminal velocities of $500$--$650 \km \s^{-1}$, depending on latitude (Smith et al. 2003).
There is no dispute about the origin of this component.

\item \emph{M-component.} Although less prominent, the M-component has a flat profile
in the velocity range $-v = 620$--$720 \km \s^{-1}$.
We are not sure about its nature, but we do see signatures that it exists at phases
11.875, 12.014 and 12.041, although less prominently.
We attribute it to parts in the primary wind that are accelerated to
somewhat higher velocities.
These may be, for example, clouds with higher opacity, e.g., due to dust formation.
We note that before the 2009 periastron passage the primary wind might have undergone some fluctuations
in its velocity, which may have been evident in the X-ray lightcurve (Kashi \& Soker 2009d).

\item \emph{F-component.} We identify the main parts of this component with the flat part in the
velocity range $-v \simeq 1150$--$1400 \km \s^{-1}$, and the tail formed by faster and lower column density gas
with absorption extending to $-v = 1900 \km \s^{-1}$.
This tail is not due to line broadening in the main part of the absorbing component.
The F-component does not exist at all at phases 11.875, 12.014 and 12.041.
Namely, it is a transient component appearing for a short time before periastron passage.
In its flat part this component absorbs $\sim 25-30$ per cent of the continuum at phase 11.998,
and $\sim 20 $ per cent at phase 11.991.
\end{enumerate}

We attribute the F-component to pre and post-shock secondary wind material close to the binary system
(Kashi \& Soker 2009b).
The post-shock gas resides in the conical shell.
We further assume that the secondary wind gas absorbs most of the continuum emitted by the
gas in the conical shell (emission that increases just before periastron passage),
in addition to absorbing some fraction of the continuum emitted by the primary wind.
It is this component that is in dispute between G2010 and us.
Although G2010 and us attribute it to the colliding winds, in the model of G2010
the absorbing gas is at a distance of tens of $\AU$ from the binary systems, while in our
model the absorbing material is relatively close to the binary system.

One thing to notice is that the first observation after the 2009 spectroscopic event is at phase 12.014,
before the system exits from the X-ray minimum (Corcoran 2010),
and before the hard ionizing radiation from the secondary resumes after exiting the dense primary wind.
As we note in section \ref{sec:problem}, this holds difficulties to the model of G2010.

\section{PROBLEMS WITH THE MODEL OF G2010}
\label{sec:problem}

The purpose of this section is to set the motivation for a model
where the fast absorbing material reside close to the secondary star,
and the secondary star is in front of the primary near periastron passage.
We critically examine a model with an opposite binary orientation, and reveal its problems.

G2010 present (in their fig. 12m-o) the total column density as a
function of gas velocity at 5 phases according to their assumptions.
Their main assumptions are that:
(1) The photospheric radius of the primary at $1.083 \mum$ is $2.2 \AU$.
(2) A point source can be used for the central source of the $1.083 \mum$ continuum emission.
(3) The central source is the only source of the continuum (below we claim that the conical shell is an additional source).
(4) G2010 did not consider even qualitatively the change in level population.
In their model the change in level population must be significant; below we argue why this is unlikely.

In more details, their fig. 12 gives the density of material as a function of line of sight
distance to the primary for 5 orbital phases (11.875, 11.991, 11.998, 12.014 and 12.041)
and 6 orientations ($\omega=0^\circ$, $50^\circ$, $90^\circ$, $180^\circ$, $243^\circ$, and $270^\circ$).
In addition, the column density along the same assumed line of sight to the primary star is given as a function of the line of sight velocity,
for each orientation and phase.
Fig. 12 of G2010 allows us to deduce the basic properties of their model.

G2010 did not relate the total column density to the absorption of the He~I~$\lambda10830 \rm{\AA}$ line.
We attempt to do so and check what the results of G2010 imply.
As we now show, their model, with a preferred orientation of $\omega=243^\circ$, suffers from severe
problems because the column density does not follow even qualitatively the behavior
of the observed absorption.
In their attempt to resolve this discrepancy, G2010 claim that the presence of high velocity gas in
our line of sight is a necessary, but not sufficient condition for the presence of high velocity absorption.
As the amount of absorption depends on the population of the lower energy level of the
He~I~$\lambda10830 \rm{\AA}$ line ($2~^3S$), they attribute the differences in absorption
to different population of the $2~^3S$ atomic level.

\subsection{Pre-periastron passage}

{}For the three phases before periastron passage (11.875, 11.991, 11.998),
G2010 model (in their fig. 12o) predicts the absorbing column density in the range
$-v = 2000$--$3000 \km \s^{-1}$ to be about equal to that in the velocity range $-v = 1600$--$1800 \km \s^{-1}$.
Observations show substantial absorption in the $-v = 1600$--$1800 \km \s^{-1}$ range
at phases 11.991 ($\sim 3-10$ per cent absorption) and 11.998 ($\sim 5-13 $ per cent absorption),
while there is no absorption at $-v > 1900 \km \s^{-1}$ (fig. 1 of G2010).
G2010 attribute the difference to stratification in the population of the
$2~^3S$ level, from which the He~I~$\lambda10830 \rm{\AA}$ line is absorbed.
However, they do not present a qualitative study of the way the population of the
$2~^3S$ level can evolve so rapidly.
We find it extremely unlikely (unless there is a carefully arranged fine tuning)
that differences in populating the $2~^3S$ level can account for this difference.
The reason is as follows.

According to G2010 model during the phase interval 11.991--11.998 the absorbing gas at
$-v = 2000 \km \s^{-1}$ resides at a distance of $r_a \simeq 3.5 a$
from the binary system, where $a \sim 16 \AU$ is the semimajor axis (their fig. 12n).
They also find that the gas absorbing in the velocity range $-v = 1600$--$1800 \km \s^{-1}$ resides at
$r_a \simeq 2.5-3 a$.
During this phase interval, the binary orbital separation is $r \la 0.2 a$.
Therefore, the differences in the radiation of the two stars at $r \simeq 2.5-3 a$ and
$r \simeq 3.5 a$ are small, and cannot account for the required large differences
in the population of the $2~^3S$ level.
Moreover, the increase in the amount of absorption as the system approaches
periastron in the model of G2010 is attributed to an increase in the population of the $2~^3S$ level.
This is probably (G2010 do not specify the reason) because the secondary gets deeper into the primary wind,
and less of its radiation reaches the absorbing gas.
It is expected that the outer regions, where gas moving at $-v = 2000 \km \s^{-1}$ resides,
will get less of the secondary radiation, and hence will absorb more.
This is opposite to what their model requires.

A potential way to handle the problem is that the material closer-in is of higher density,
and it recombines first or its collisional excitation is more efficient.
This would not work either.
The rate of these two processes is linear with density.
The density ratio between the region absorbing in the $-v = 1600$--$1800 \km \s^{-1}$
range to that absorbing in the $-v \simeq 2000 \km \s^{-1}$ range is $\sim 2$ (G2010; fig. 12m,n).
Therefore, the absorption at $-v \simeq 2000 \km \s^{-1}$ is expected to be
one half of that in the $-v = 1600$--$1800 \km \s^{-1}$ velocity range.
This amounts to $\sim 1.5-5 $ per cent of the flux at phase 11.991 and
$\sim 2.5-6.5 $ per cent of the flux at phase 11.998.
This is clearly ruled out by observations, as the absorption at $-v = 2000 \km \s^{-1}$
observed in these two phases is practically zero.

\subsection{Post-periastron passage}

In phase 12.014 (28 days after phase 12) the column density calculated by G2010
in the velocity range $-v = 1200$--$1400 \km \s^{-1}$ is a factor of 2-3 lower than that
at phase 11.998, while in $-v = 1600$--$1800 \km \s^{-1}$ this ratio is $<2$ (their fig. 12o).
At phase 12.014 the secondary radiation has not started yet to ionize the material at large distances,
as it is still inside the dense region of the primary wind.
We know this from the behavior of the highly ionized lines, e.g.,
in the Weigelt blobs, which show that the hard radiation resumed (in the 2003.5 spectroscopic event)
well after the X-ray minimum ended (Damineli et al. 2008b; Mehner et al. 2010).
In the 2009 spectroscopic event the X-ray minimum ended at phase $\sim 12.014$ (Corcoran 2010),
when one of the He~I~$\lambda10830 \rm{\AA}$ line measurements was conducted.
Namely, at this phase the secondary hard ionizing radiation did not ionize the He
in the relevant distance of tens of $\AU$.
The density in the relevant region at phase 12.014 is not much different from that at phase 11.998.
If it was for recombination, there should have been more time for the gas at phase 12.014 to recombine.

Over all in the velocity ranges $-v = 1200$--$1400 \km \s^{-1}$ and $-v = 1600$--$1800 \km \s^{-1}$
the column density calculated by G2010 at phase 12.014 is
$\sim 0.3-1$ that at phase 11.998, and the ionization state should be the same.
Therefore, it is expected that the absorption at phase 12.014 in the velocity ranges
$-v = 1200$--$1400 \km \s^{-1}$ and $-v = 1600$--$1800 \km \s^{-1}$ would be $\sim 0.3-1$ times
that at phase 11.998, or even above its value at phase 11.998.
This is contrary to observations, as there is no absorption at all for
$-v > 1100 \km \s^{-1}$ at phase 12.014 (fig. 1 of G2010).
This is a severe problem for the model presented by G2010.

Not only the model of G2010, but any other model that would assume absorption by materiel residing at a large distance
(a few $\times 10a$ or more) is problematic.
The reason is that the absorption of the He~I~$\lambda10830 \rm{\AA}$ line in high velocities is
a transient event, which occurs very close to periastron, and must be related to variations in the
system shortly before periastron.
The material far from the binary system does not suppose to show fast variations close to periastron,
and for that it is a bad candidate for an absorber of the He~I~$\lambda10830 \rm{\AA}$ line.

\section{THE MODEL}
\label{sec:omega90}

\subsection{The column density of the absorbing gas}
\label{subsec:omega90:absorbing}

We perform hydrodynamic simulations of the colliding winds to show that the column density of the conical shell and
the pre-shocked secondary wind at high velocities is high enough to account
for the absorption of the He~I~$\lambda10830 \rm{\AA}$ line.
The simulations are performed with Virginia Hydrodynamics-I (VH-1), a high
resolution multidimensional astrophysical hydrodynamics code developed by
J. Blondin and collaborators (Blondin et al. 1990; Stevens et al., 1992; Blondin 1994).
The code includes radiative cooling at all temperatures of $T > 2 \times 10^4 \K$.
The radiative cooling is set to zero for temperatures of $T<2 \times 10^4 \K$, and the gas cannot cool to lower temperatures.
The cooling function $\Lambda (T)$ (for solar abundances) is taken from Sutherland \& Dopita (1993; their table 6).

\begin{figure}
\includegraphics[width=1\columnwidth]{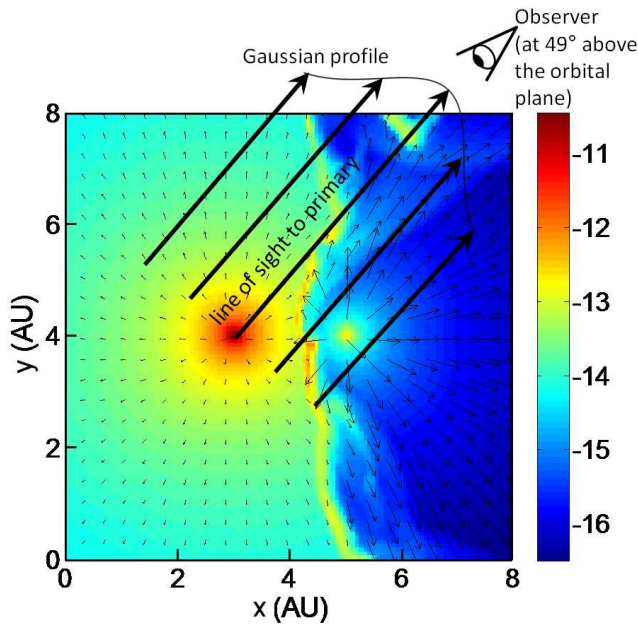}
\caption{
Our hydrodynamical simulations showing density and velocity maps for an orbital separation of $r = 2 \AU$.
The bar on the right gives the density color-code in units of
$\log[\rho~(\rm{g~cm^{-3}})]$.
The thin arrows represent the velocity in the (x,y) plane (the orbital plane)
and are scaled with the axes such that one unit represents $5000 \km \s^{-1}$.
The secondary is the star on the right at $(x,y)=(5,4) \AU$.
The arrows are in the line of sight direction to an observer at $(\omega,i)=(90^\circ,41^\circ)$ and at phase 0.9964.
The column density from each line of sight is weighted with a Gaussian (equation \ref{eq:Gaussian}).
}
\label{fig:simulation}
\end{figure}

The numerical simulations were performed in the Cartesian geometry $(x,y,z)$ mode
of the code (a 3D calculation), where the orbital plane is the $(x,y)$ plane with the $x$ axis taken to be parallel
to the line connecting the two stars.
Our numerical simulation grid has 112 equal-size grid points along each axis.
The primary and the secondary are located in the middle of the $y$ and $z$ axes.
The binary separation between the two stars is set to be constant at each numerical run
(for the justification see Akashi \& Soker 2010).
The mass loss rates and velocities of the winds are
$\dot M_1= 3 \times 10^{-4} ~\rm{M_\odot} \yr^{-1}$ and
$\dot M_2= 10^{-5} ~\rm{M_\odot} \yr^{-1}$, and $v_1 = 500 \km \s^{-1}$
and $v_2 = 3000 \km \s^{-1}$, respectively (e.g., Pittard \& Corcoran 2002;
Kashi \& Soker 2008b and references therein).
We use eccentricity of $e=0.9$ and semi-major axis of $a=16.64 \AU$.
We performed four runs with binary separation of $r=$ $2$, $3$, $4$, and $10$ $\AU$,
and in each created a density profile and wind velocity field around the two stars.
At the location of each star we inject its appropriate wind.
We let the flow reach an approximately steady state.
An example of the results of the simulation is shown in Fig. \ref{fig:simulation}.

For each run we calculate the direction to the observer for a system with $(\omega,i)=(90^\circ,41^\circ$),
and for the phase before periastron corresponding to the binary separation of the run.
We take a plane perpendicular to the line of sight and located behind the primary,
and from this plane we calculate the column density along
a grid of lines of sight to the observer.
We select the location of this plane in an approximately optimal way to include as much of the grid as possible.
This procedure favors accuracy in the calculation of the column density of the high velocity gas towards the observer.
Each line of sight is divided into cells, as the number of $z$ planes the line of sight crosses in our grid.
We calculate the column density along each line of sight as function
of the velocity projected along the line of sight.
Like G2010 we use velocity bins of $50 \km \s^{-1}$.

We adopt the conclusions of G2010 that the main source of the $1.083 \mum$ continuum emission comes mostly
from a region of size $\sim 2.2 \AU$ centered on the primary star.
We term it the central source.
Though G2010 used a point source for the central source of the $1.083 \mum$ continuum emission,
they mentioned that it is of size of several $\AU$,
based on observations by Weigelt et al. (2007) in the K-band,
which were calibrated to $1.083 \mum$ (which is within the I-band),
with the help of the theoretical results of Hillier et al. (2001).
Weigelt et al. (2007) found the decrease function of the central source intensity with projected distance
from the center $r_p$ to be close to a Gaussian (their fig. 6).
We therefore take the emission to be from a radius of $R_{\rm{em}}=4.4\AU$,
and weighted the intensity as a Gaussian with $\sigma=2.2\AU$
\begin{equation}
I = I_0 \exp(-r_p^2/2 \sigma^2) ; \rm{~for~} r_p \leq R_{\rm{em}} ,
\label{eq:Gaussian}
\end{equation}
where $I_0$ is the intensity at the center (the cut-off at $R_{\rm{em}}$ is used due to numerical constrains
and results in a negligible error).

\begin{figure}
\includegraphics[width=1\columnwidth]{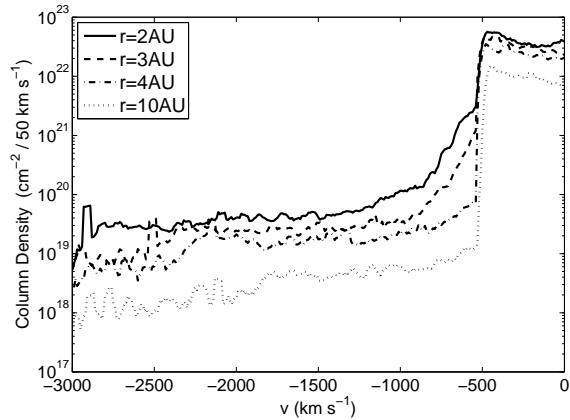}
\caption{
The calculated column density as a function of projected velocity along the line of sight.
}
\label{fig:N_H_vs_v}
\end{figure}

The column density from each line of sight is weighted with the Gaussian in equation (\ref{eq:Gaussian})
according to its distance $r_p$ from the line of sight of the primary (Fig. \ref{fig:simulation}).
This procedure gives us an effective column density in each velocity interval and at each phase.
The results are presented in Fig. \ref{fig:N_H_vs_v}.
%

{}From Fig. \ref{fig:N_H_vs_v} we learn the following:
(1) For $\omega=90^\circ$ the column density in the conical shell and the undisturbed wind is high enough to account for
the absorption of the He~I~$\lambda10830 \rm{\AA}$ line for our favored orientation of $\omega=90^\circ$.
As the model of G2010, our model is unable to calculate the population of the $2~^3S$ level of He~I,
and hence we cannot calculate the exact absorption in the He~I~$\lambda10830 \rm{\AA}$ line.
(2) The high column density for $-v \gtrsim 1000 \km \s^{-1}$ is a transient component which appears only for a short period
before periastron passage.
(3) The column density clearly increases as the system approaches periastron (note that in fig. 12 of G2010 this is not the case).
(4) The column density decreases towards higher blue velocities, despite of minor fluctuations due to the limited numerical resolution.
(5) At short binary separations ($r \lesssim 4 \AU$; $t \gtrsim 0.987$) there is a plateau in the column density between
$\sim 1400 \km \s^{-1}$ and $\sim 2100 \km \s^{-1}$, followed by a decrease in column density towards higher blue velocities.

Our model gives that at very high velocities of $-v > 1900 \km \s^{-1}$ the column density is also large,
namely we have more absorbing gas than required to account for the observations.
This discrepancy may be the result of an anisotropic secondary wind velocity close to periastron passage,
that can occur if the acceleration of the secondary wind is not isotropic,
an effect which was not taken into account in our simulation.
Such an effect can reduce the amount of gas which outflows at $-v > 1900 \km \s^{-1}$.
The acceleration can be efficient along polar directions, but below
$i \simeq 60^\circ$ radiation from the primary and previous accretion of blobs might cause the secondary wind to be slower.
This explanation cannot work for correcting the surplus absorbing gas in the model of G2010,
as their absorbing material is much further away from the colliding winds region.

\subsection{Covering the emission source with the conical shell}
\label{subsec:omega90:conical shell}

{}From Fernandez-Lajus et al. (2010) we find that the I-band continuum has a quiescence value of $I=3.55$ mag.
We sample the increase in the I-band continuum flux above the quiescence value, $f_c(t)$,
from the observations of Fernandez-Lajus et al. (2010).
These observations show that close to the 2009 periastron passage the I-band continuum flux
(and therefore the $1.083 \mum$ line flux) has increased by $f_c(t=0) \simeq 20$ per cent
above the quiescence value (Fig. \ref{fig:Fobs}).
We attribute most of this extra emission to the conical shell close to the binary system.
We further assume that there is a strong absorption by the conical shell
and the pre-shocked secondary wind of the He~I~$\lambda 10830 \rm{\AA}$ line.

We analytically calculate the fraction of the central source of the $1.083 \mum$ continuum emission region that is covered by
the conical shell (namely, covered behind the conical shell from the line of sight of the observer).
The parameters of the stars and the conical shell are taken as in our previous papers (Kashi \& Soker 2009b, c, and references therein).
The contact discontinuity is approximated as an hyperboloid with semi-opening angle $\phi_a \simeq 60^\circ$.
This is an approximation, as numerical simulations
show that the structure is much more complicated (Akashi \& Soker 2010; Parkin et al. 2011).
We take into account the rotation of the conical shell close to periastron,
caused by the non negligible orbital velocity relative to the primary wind velocity,
and the acceleration of the primary wind (Kashi \& Soker 2009b, c).
The eccentricity is taken to be $e = 0.9$, such that the orbital separation at periastron is $a_p = 1.66 \AU$.
This implies that the secondary star and part of the conical shell are inside the central $1.083 \mum$ continuum
emission region near periastron passage.
Our assumptions break down after periastron passage, when the conical shell collapses.

\begin{figure}
\includegraphics[width=1\columnwidth]{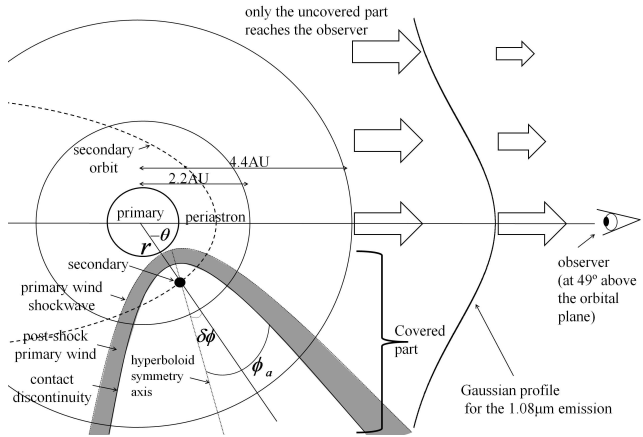}
\caption{
The geometry of the binary system and the $1.083 \mum$ continuum emission region.
}
\label{fig:geometry}
\end{figure}

\begin{figure}
\includegraphics[width=1\columnwidth]{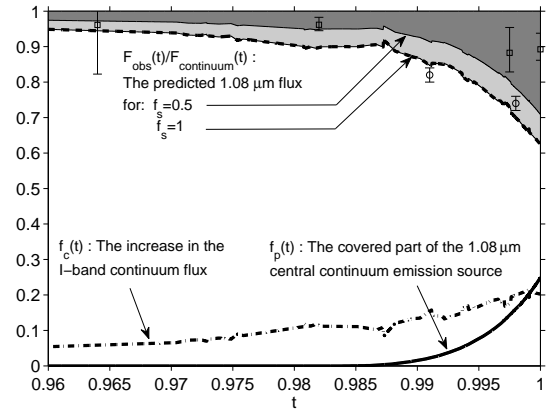}
\caption{
Behavior of several properties just before periastron passage.
The I-band continuum fluxes before the 2009.0 and 2003.5 periastron passages are very similar.
Circles and squares :The observed decrease in the flat part of the F-component prior to the 2009.0 and 2003.5 periastron passages, respectively.
The error bars are calculated from the maximum and minimum absorption values in the velocity range of the F-component.
}
\label{fig:Fobs} \end{figure}

We again use our preferred orientation $(\omega,i)=(90^\circ,41^\circ$)
The geometry of the calculation is shown in Fig. \ref{fig:geometry}.
For every orbital phase we divide the observed half sphere around the primary that is closer to the observer into spherical
volume elements (bins).
For each emission source element we check whether the conical shell is between the observer and the emission source element,
or whether it is within the conical shell itself.
If the conical shell is between the observer and the emission source element,
we use our assumption of high optical depth in the conical shell,
and assume the emission from the covered source is totally absorbed.
In other words, if the emission source element is either of the two (behind the conical shell or inside it),
we consider this element to contribute to the $1.083 \mum$ line absorption.
The contribution of each element is weighed as the Gaussian, with the projected distance to the line of sight, as discussed above.

The calculated flux of the $1.083 \mum$ flux line according to our model is
\begin{equation}
\frac{F_{\rm{obs}}(t)}{F_{\rm{continuum}}(t)} = \frac{1-f_p(t)+f_c(t)(1-f_s)}{1+f_c(t)} ; \quad {\rm F-component} \label{eq:Fobs}
\end{equation}
where $F_{\rm{continuum}}(t)$ is the $1.083 \mum$ continuum flux (given by Groh et al. 2010, as defined above),
$f_p$ is the fraction of the central continuum source covered by the conical shell,
$f_c$ is the increase in the I-band continuum flux above the quiescence value (Fernandez-Lajus et al. 2010),
and $f_s$ is the fraction of the conical shell emission that is absorbed by the F-component.
If the conical shell self absorb all its emission, then $f_s=1$,
while if it absorbs only from its volume that is behind the conical shell,then $f_s=0.5$.
We present results for both these values that bound our expectation.
Fig. \ref{fig:Fobs} presents $F_{\rm{obs}}(t)/F_{\rm{continuum}}(t)$ close to periastron, as well as $f_c$ and $f_p$.
The conical shell absorbs up to $f_p(t=0) \simeq 25$ per cent of the intensity of the
$1.083 \mum$ continuum emission from the central source (the primary star).
When $\omega = 90^\circ$ our model accounts for the fast increase in absorption of the flat part (F-component; Fig. \ref{fig:profile}).
Considering the many uncertainties and the simplicity of our model,
our results are in reasonable agreement with the observed decrease
in the flat part of the F-component in the velocity range $-v \simeq 1150$--$1400 \km \s^{-1}$
(the tail to $-v = 1900 \km \s^{-1}$ is part of this component, and behaves in the same way).

The sizes of the central and extra emission sources are important for this modelling.
G2010 use a point-like continuum source in their quantitative calculation of the column density of the absorbing material
(total column density, not only that of the relevant atoms).
For the model of G2010, where $\omega = 243^\circ$, it does not make much difference to use a point source,
as their absorbing material is several tens of $\AU$ from the source.
However, calculating the column density to a point source fails to reproduce results of a model where the absorption occurs
a few $\AU$ from the stars, such as  our model (Kashi \& Soker 2009b), which we improve here.
G2010 do discuss (in their section 5.3) a $2.2 \AU$ extended source for an orientation of $\omega = 90^\circ$,
and find the duration of the high velocity absorption to be too short.
Our point here is that an extra source must be included to take into account the increase in the I-band and for the extra absorption.
In our model the extra source is the conical shell.
Our results show that one must use an extended emission source, both for the central source (the primary star)
and the extra source from the conical shell.
Using the two extended sources overcomes the objection of G2010 to the $\omega = 90^\circ$ orientation.

\section{SUMMARY AND DISCUSSION}
\label{sec:summary}

The He~I~$\lambda10830 \rm{\AA}$ line of $\eta$ Car has been observed across the 2009 periastron passage by G2010.
We identify three main absorbing components in the He~I~$\lambda 10830 \rm{\AA}$ line profile (Fig. \ref{fig:profile}).
Two components can be identified with the dense primary stellar wind;
the P-component ($500$--$650 \km \s^{-1}$) and the M-component ($-v = 620$--$720 \km \s^{-1}$).
The third, more interesting F-component shows a flat part in the
velocity range $-v \simeq 1150$--$1400 \km \s^{-1}$, with a tail to $-v = 1900 \km \s^{-1}$.
The tail results from faster moving gas, and not from line broadening.
There is no absorption in this line at $-v > 1900 \km \s^{-1}$.

G2010 present their interpretations of their observations,
concluding that the orientation of the binary system is such that $\omega=243^\circ$
(the primary closer to the observer just before periastron).
In section \ref{sec:problem} we point out some problems in the model of G2010,
in particular that they cannot account for the transient nature of the line.
The model of G2010 fails to reproduce, even qualitatively, the observed absorption profiles
of the He~I~$\lambda10830 \rm{\AA}$ line, and cannot account for the absence of the F-component after periastron passage.
In analyzing our (Kashi \& Soker 2008b, 2009b) preferred orientation of $\omega=90^\circ$,
G2010 use a point-like central continuum source for calculating the column density of the absorbing material.
Based on that, they claim that the $\omega=90^\circ$ orientation is not compatible with their observations.
In section \ref{sec:omega90} we find that the usage of a central point source
does not do justice to the $\omega=90^\circ$ case.
This is the main reason why G2010 fail to reproduce results of a
model where the absorption occurs a few $\AU$ from the stars, like in our model.

We improve our pervious model (Kashi \& Soker 2009b) which assumed that the conical shell
is the main absorber of the high velocity component of the He~I~$\lambda10830 \rm{\AA}$ line (section \ref{sec:omega90}).
Our model and results can be summarized as follows:
\begin{enumerate}
\item We assume that the central star is a Gaussian weighted extended continuum source as given by equation (\ref{eq:Gaussian}).
This assumption stands on a solid ground (G2010; Weigelt et al. 2007).

\item We assume that the increase in the $1.083 \mum$ continuum near periastron passage comes from
the conical shell (the collision region of the two winds).
The main source is the shocked primary wind.
This is based on analysis we performed in previous papers.

\item We assume that the main absorbing gas of the He~I~$\lambda10830 \rm{\AA}$ line
is both pre and post-shock material in the fast secondary wind.
This assumption stands on a solid ground.
Based on recent numerical simulations (Akashi \& Soker 2010), we showed in section \ref{sec:omega90}
that the column density of the fast moving gas is as high as found by G2010.
Some segments of the wind are at a temperature of $\sim 10^4 \K$, and there are enough recombining He atoms.
This will be calculated in a future paper based on high resolution numerical simulations.

\item We concentrate on the absorption of the F-component.
We assume that in the flat velocity range of the F-component (see Fig. \ref{fig:profile}),
the absorbing gas in the conical shell absorbs most of the conical shell emission.
We therefore consider two limiting values for this parameter $f_s = 0.5$ and $f_s = 1$.

\item We assume that the optical depth of the conical shell in the flat
velocity range of the F-component is very high, such that it absorbs
all the radiation of the central source it hides from our line
of sight (in the flat part).

\item We assume that the colliding winds shell has an hyperbolic shape.
We consider its tilt due to the orbital motion.
In calculating the shape and tilt the primary wind acceleration zone is considered.

\item We assume, based on our previous papers (Kashi \& Soker 2008b,2009c), that the secondary
is toward us near periastron ($\omega =90^\circ$).
We take an inclination angle of $i=41^\circ$, and the other commonly used binary parameters
(semimajor axis, eccentricity, etc.).

\item We note that after periastron passage some of the assumptions break down,
because accretion is likely to occur (Kashi \& Soker 2009a; Akashi \& Soker 2010).

\item We calculate the part of the central source covered by the conical shell, for $\omega=90^\circ$.

\item We calculate the covered part of the $1.083 \mum$ central continuum emission source (equation \ref{eq:Fobs}).
By that we show that for $\omega=90^\circ$ our model explains the observed increase in absorption strength of the
F-component close to periastron passage (Fig. \ref{fig:Fobs}), and explains its transient nature.

\item We therefore conclude that the orientation is indeed that the secondary is toward us near periastron ($\omega =90^\circ$).

\item In addition, using hydrodynamic simulations (Fig. \ref{fig:simulation} and \ref{fig:N_H_vs_v})
we find that the He~I in the conical shell and in the pre-shocked secondary wind
has a substantial column density in the velocity range $-v = 1150$--$1400 \km \s^{-1}$.
We assume that the fraction (out of total He atoms and ions) of the He~I in the $2~^3S$ level is high enough to absorb most radiation in the
$-v < 1400 \km \s^{-1}$ (with a tail up to $-v \simeq 1900 \km \s^{-1}$).
\end{enumerate}

The finding that the secondary is closer to us near periastron requires that some properties
near periastron passage be explained by the accretion of the primary wind onto the secondary star (Kashi \& Soker 2008a, 2009d).
For accretion to occur onto the secondary star, that has a strong wind of his own,
the binary must be close (Akashi \& Soker 2010) and to interact strongly with the primary.
The accretion that occurs near periastron passage is crucial to the understanding not only the present
behavior of $\eta$ Car, but also its behavior during the major eruptions it undergone in the 19th century,
the Great Eruption (GE) and Lesser Eruption.

The debate on the absorbing source of the blue wing has implications far beyond the specific question
on the orientation of the major axis of the binary system.
The essence of the debate is the nature of the binary interaction process, which
has implications on the nature of LBV major eruptions, and mass loss by very massive stars
(e.g., Soker 2001, 2005, 2007; Kashi \& Soker 2010; Smith et al. 2010; Smith 2010a,b).

The GE is the best studied example for a major LBV eruption, and serves as a test
case for theories and models.
A high rate accretion during the GE could have supplied the extra luminosity for
20 years, and the accreting secondary star could have launched two jets that shaped
the bipolar structure -- the Homunculus (Soker 2001).
Most likely other LBV major eruptions are
also related to binary interaction (Kashi \& Soker 2010), as was argued for the
17th century eruption of P~Cygni (Kashi 2010), rather than a single star phenomena as suggested by, e.g., Smith (2007).
Smith \& Owocki (2006) suggested that LBVs lose most of their envelope mass during major eruptions.
Therefore, the presence of a strongly interacting companion can play a major role in the evolution
of very massive stars.
As suggested by Kashi et al. (2010), major mass transfer events in LBVs are related to optical transient objects and
have a common powering mechanism -- accretion onto a companion star.
Understanding the binary interaction in $\eta$ Car will shed light on other objects where binary interaction
is thought to shape circumstellar nebulae, like planetary nebulae, symbiotic systems, and
related objects such as the Red Rectangle.

\section*{acknowledgments}
We thank Jose H. Groh for very valuable and enlightening comments,
and an anonymous referee for comments that helped to improve the paper.
We acknowledge Eduardo Fernandez-Lajus for using his I-band observations,
and Augusto Damineli for using his observations of the He~I~$\lambda10830 \rm{\AA}$
line across the 2003.5 periastron passage.
This research was supported by the Asher Fund for Space Research at the Technion
and a grant from the Israel Science Foundation.

\appendix

\textbf{APPENDICES (ASTRO-PH VERSION ONLY)}

\section{THE BRIEF ABSORBTION IN THE CONICAL SHELL}
\label{sec:absorb_shell}

A support to a short duration absorption by the conical shell near periastron
passage might come from the behavior of the Hydrogen H$\delta$ line.
The Hydrogen atom is a simple atom, lacking the complications
resulting from absorption from a meta-stable state as for the
He~I~$\lambda10830 \rm{\AA}$ line.
The complication is in that it is not simple to estimate the
population of the relevant atomic state.
Nielsen et al. (2009) present high spectral resolution ($R \sim 80,000$)
of the hydrogen H$\delta$ line, as well as of the
He~I~$\lambda7067 \rm{\AA}$ line, across the 2003.5 spectroscopic event.
The He~I~$\lambda7067 \rm{\AA}$ line shows Doppler shift variations with the orbital phase that
fit emission and absorption in the acceleration zone of the secondary wind (Kashi \& Soker 2007, 2008b).
The variations in the Doppler shift of the H$\delta$ line are much smaller, perhaps
$\sim 0.3$ times those of the He~I~$\lambda7067 \rm{\AA}$ line (hard to tell).
This may suggest that part of the absorption in the H$\delta$ line is from the conical shell,
in addition to an absorption component associated with the primary wind.
Close to periastron (phase 1.001 in figure 5 of Nielsen et al. 2009) the H$\delta$ line
shows an anomalous wide blue absorption wing, reaching velocities up to $-v \simeq 900 \km \s^{-1}$.
At the closest observation times, phases 0.985 and 1.115, this broad absorption does not exist.
Though the blue wing is of slower velocities than observed in the He~I~$\lambda10830 \rm{\AA}$ line,
a resemblance between the two is very likely.
Both lines show a high velocity absorption component just before periastron, which later disappears.
Like the He~I~$\lambda10830 \rm{\AA}$ line, it is
possible that the H$\delta$ absorption line originates in the conical shell, but
in regions with higher density and lower velocity.
We note that qualitatively similar blue absorption wing close to periastron passage
also appears in the H$\alpha$ and Paschen~$\eta$ lines, although less prominent
(Davidson et al. 2005; for the H$\alpha$ this trend is better seen
in the animation prepared from these observations).
Note that in the last periastron passage of 2009 the H$\alpha$ (observed by
Richardson et al. 2010) blue wing absorption component is not prominent, but it may exist
and be difficult to see.

\section{ABSORPTION IN A JET?}
\label{sec:jet}

We here check whether an alternative absorbing material for the He~I~$\lambda10830\AA$
can be a collimated fast wind or a jet emitted from the secondary.
Behar et al. (2007) analyzed \emph{CHANDRA} MEG/HEG observations in the $3.7-7.5 \AA$ band,
where high excitation S, Si and Ar lines are observed.
Far from periastron these lines show a symmetric profile with no Doppler shift.
Close to periastron the centroid moves to $\sim -200 \rm{km~s^{-1}}$, and a wide blue wing is developed,
reaching velocities of up to $\sim -2000 \rm{km~s^{-1}}$.
Behar et al. (2007) interpreted the centroid motion to the orbital motion of the secondary,
taking $\omega=90^\circ$.
We note that though the observed velocity at phase $-0.006$ was out of their Doppler lines (their fig. 4),
taking updated inclination angle of $i=41^\circ$ instead of $i=53^\circ$ as used by Behar et al. (2007)
makes their fit much better and strengthen their claim for the emission source for the centroids.
The high velocity blue wing seen in these lines was attributed by Behar et al. (2007)
to a collimated fast wind or a jet launched by the secondary.

The wide blue wing at high velocities is clearly seen at phases $-0.028 (11.972)$ and $-0.006 (11.994)$ in fig. 3 of Behar et al. (2007),
where the mean velocity profiles constructed from nine different spectral lines from five observations are plotted.
The flux is stronger at phase $-0.028$ for velocities $-v \lesssim 1400 \rm{km s^{-1}}$.
Fig. 3 of Behar et al. (2007) presents normalized observation.
{}From the original data presented in Henley et al. (2008) we learn that the true flux at phase $-0.028$ is $\sim 1.7$
times as strong as at phase $-0.006$ so that the wide blue wing at phase $-0.028$ is even more pronounced.
Correcting for the true, not normalized, flux, we find that the wide blue wing at phase $-0.028$ is stronger than in phase $-0.006$ for
not only for $-v \lesssim 1400 \rm{km s^{-1}}$ but up to $-v \lesssim 2000 \rm{km s^{-1}}$.
In this context, there is some difficulty in adopting the jet explanation for He~I~$\lambda10830\AA$ line absorption at high blue velocities.
At phase $-0.028$, $57$ days before periastron, the accretion process have not yet started (Kashi \& Soker 2009a),
and hence an accretion disk around the secondary which is essential for producing jets does not exist.
This is also supported by the X-ray flux which is strong and raising at phase $-0.028$,
while weaker and decreasing at phase $-0.006$ (Corcoran 2010),
suggesting that substantial accretion has not yet started according to the accretion model (Soker 2005; Kashi \& Soker 2009a).
We therefore regard the explanation of a fast wind from a jet as the absorbing material for the He~I~$\lambda10830\AA$ unlikely.

It is possible that the gas emitting the highly ionized wide S, Ar, and  and Si lines and the He~I~$\lambda10830\AA$
absorption do have a common origin.
The emission and absorption occur in different places along the outflow.
It is therefore possible that the X-ray emitting gas is not a jet, but rather the secondary wind (pre and post-shock) itself.

\label{lastpage}
\end{document}